\documentclass[11pt]{article}
\usepackage{times}
\usepackage{mathfont}

\def\log{\mathop{\rm log}}

\input epsf
\def\efig#1#2{\hbox{\epsfxsize=#1\epsfbox{#2}}}

\newtheorem{theorem}{Theorem}
\newtheorem{lemma}{Lemma}

\DeclareSymbolFont{lasy}{U}{lasy}{m}{n}
\let\Box\undefined
\DeclareMathSymbol\Box{0}{lasy}{"32}
\newcommand{\qed}{\hfill$\Box$}
\newenvironment{proof}{\noindent{\bf Proof:}}{\qed\medskip}

\makeatletter

\long\def\@makecaption#1#2{
   \vskip 10pt 
   \setbox\@tempboxa\hbox{{\small #1. #2}}
   \ifdim \wd\@tempboxa >\hsize   
       {\small #1. #2}\par        
     \else                        
       \hbox to\hsize{\hfil\box\@tempboxa\hfil}  
   \fi}

\def\@begintheorem#1#2{\it\trivlist
  \item[\hskip\labelsep{\bf #1\ #2.\ }]}
\def\@opargbegintheorem#1#2#3{\it\trivlist
  \item[\hskip\labelsep{\bf #1\ #2\ {\rm(#3)}.}]}

\makeatother

\begin{document}

\title{Beta-Skeletons Have Unbounded Dilation}

\author{David Eppstein\thanks{
Dept. of Information and Computer Science,
University of California, Irvine, CA 92697-3425,
eppstein@ics.uci.edu,
http://www.ics.uci.edu/$\sim$eppstein/}}
\maketitle

\begin{abstract}
A fractal construction shows that, for any $\beta>0$,
the $\beta$-skeleton of a point set can have arbitrarily large dilation.
In particular this applies to the Gabriel graph.
\end{abstract} 

\section{Introduction}

A number of authors have studied questions of the {\em dilation}
of various geometric graphs, defined as the maximum ratio between
shortest path length and Euclidean distance.

For instance, Chew~\cite{C86} showed
that the rectilinear Delaunay triangulation has dilation at most
$\sqrt{10}$ and that by placing points
around the unit circle, one could find examples for which the Euclidean
Delaunay triangulation has dilation arbitrarily close to $\pi/2$.
In the journal version of his paper~\cite{C89},
Chew added a further result, that the graph obtained by Delaunay
triangulation for a convex distance function based on an equilateral triangle
has dilation at most $2$.
Chew's conjecture that the Euclidean Delaunay dilation was constant was
proved by Dobkin et al.~\cite{DFS90}, who showed that the Delaunay
triangulation has dilation at most $\varphi\pi$ where $\varphi$ is the
golden ratio $(1+\sqrt{5})/2$.  Keil and Gutwin~\cite{KG89} further
improved this bound to ${2\pi\over3\cos(\pi/6)} \approx 2.42$.

\begin{figure}[t]
$$\efig{2in}{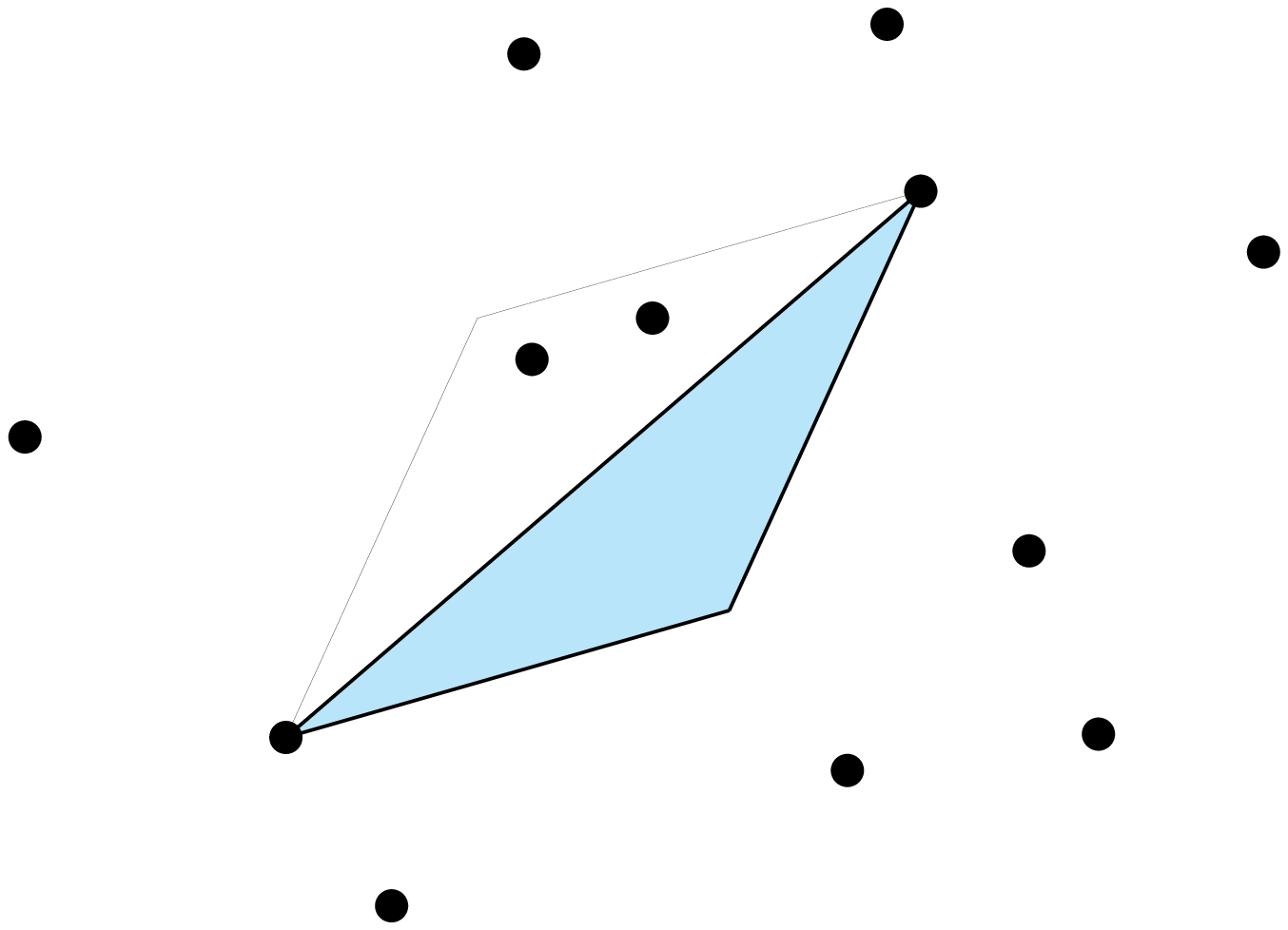}
\qquad\qquad
\efig{1.5in}{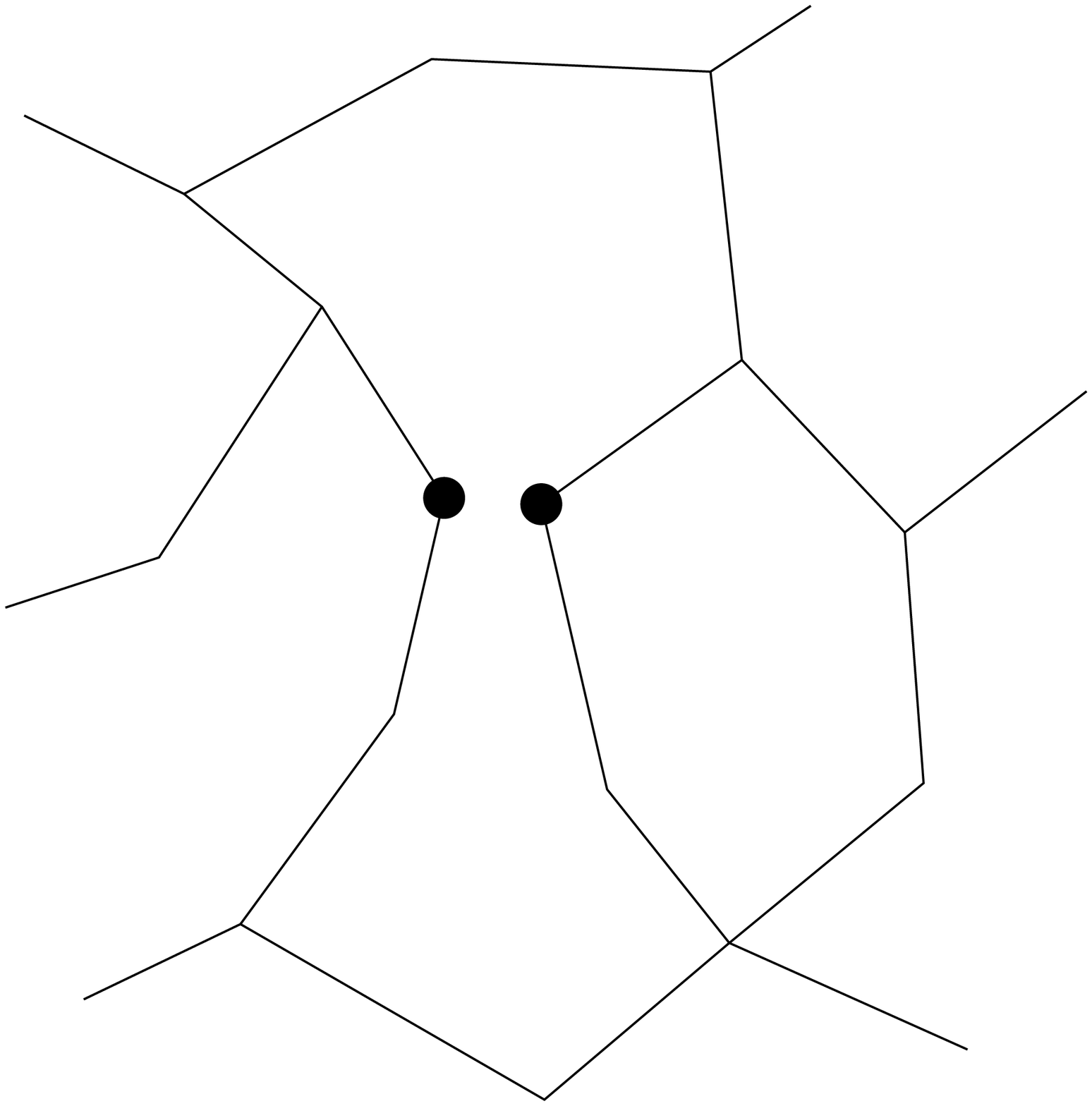}$$
\caption{(a) Diamond property: one of two isosceles triangles on edge is empty.
(b) Graph violating good polygon property: ratio of diagonal to boundary
path is high.}
\label{F:dj}
\end{figure}

Das and Joseph~\cite{DJ89} showed that these constant dilation bounds
hold for a wide variety of planar
graph construction algorithms, satisfying the following two simple conditions:
\begin{itemize}
\item {\bf Diamond property}.  There is some angle $\alpha<\pi$,
such that for any edge $e$ in a graph constructed by the algorithm,
one of the two isosceles triangles with $e$ as a base and with apex
angle $\alpha$ contains no other site.
This property gets its name because the two triangles together form a
diamond shape, depicted in Figure~\ref{F:dj}(a).
\item {\bf Good polygon property.} There is some constant $d$ such that
for each face $f$ of a graph constructed
by the algorithm, and any two sites $u$, $v$ that are visible to each
other across the face, one of the two paths around $f$ from $u$ to $v$ has
dilation at most~$d$.
Figure~\ref{F:dj}(b) depicts a graph violating the good polygon property.
\end{itemize}

Intuitively, if one tries to connect two vertices by a path in a graph that
passes near the straight line segment between the two,
there are two natural types of obstacle one encounters.
The line segment one is following may cross an edge of the graph,
or a face of the graph; in either case the path must go around these obstacles.
The two properties above
imply that neither type of detour can force the dilation of the pair of
vertices to be high.

For a survey of further results on dilation, see~\cite{E}.
Our interest here is in another geometric graph, the {\em
$\beta$-skeletons}~\cite{KR85,V92},
which have been of recent interest for their use in
finding edges guaranteed to take part in the minimum weight
triangulation~\cite{CX95,K94,Y95}
As a special case, $\beta=1$ gives the {\em Gabriel graph},
a subgraph of the Delaunay triangulation and
the relative neighborhood graph, and a supergraph of the minimum
spanning tree.
These graphs have a definition (given below)
closely related to Das and Joseph's diamond property.
The value $\beta$ is a parameter that can be taken arbitrarily close to zero;
for any point set, as beta approaches zero, more and more edges are added
to the $\beta$-skeleton until eventually one forms the complete graph.
Therefore it seems reasonable to guess that, for sufficiently
small $\beta$, the $\beta$-skeleton should have bounded dilation.
Such a result would also fit well with Kirkpatrick and Radke's
motivation for introducing $\beta$-skeletons in the study
of ``empirical networks'': problems such as modeling the
probability of the existence of a road
between cities~\cite{KR85}.

In this paper, we show that this is surprisingly not the case.
For any $\beta$, we find point sets for which the $\beta$-skeleton
has arbitrarily high dilation.  Our construction uses fractal curves
closely related to the Koch snowflake.  We show that the point set
can be chosen in such a way that the $\beta$-skeleton forms a path
with this fractal shape; the fact that the curve has a fractal dimension
greater than one then implies that the graph shortest path between its
endpoints has unbounded length.

\section{Beta-skeletons}

The $\beta$-skeleton~\cite{KR85,V92} of a set of points is a graph, defined to contain
exactly those edges $ab$ such that no point $c$ forms an angle $acb$ greater
than $\sin^{-1} 1/\beta$ (if $\beta>1$) or $\pi-\sin^{-1}\beta$ (if $\beta<1$).

\begin{figure}
$$\efig{3in}{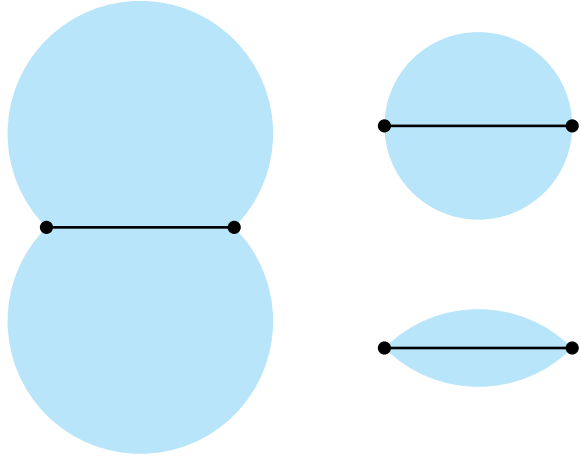}$$
\caption{Empty regions for $\protect\sqrt 2$-skeleton, Gabriel graph,
and $1/\protect\sqrt 2$-skeleton.}
\label{F:er}
\end{figure}

Equivalently, if $\beta > 1$, the $\beta$-skeleton can be defined
in terms of the union $U$ of two circles, each having $ab$ as a chord
and having diameter $\beta\,d(a,b)$.
Edge $ab$ is included in this graph exactly when $U$ contains no points
other than $a$ and $b$.

If $\beta = 1$, an edge $ab$ is included in the $\beta$-skeleton
exactly when the circle having $ab$ as diameter contains no points
other than $a$ and $b$.  The 1-skeleton is also known as the
{\em Gabriel graph}~\cite{GS69}.

If $0 < \beta < 1$, there is a similar definition in terms of the intersection
$I$ of two circles, each having $ab$ as a chord and
having diameter $d(a,b)/\beta$.
Edge $ab$ is included in the $\beta$-skeleton exactly when $I$ contains no points
other than $a$ and $b$.

Figure~\ref{F:er} depicts these regions for $\beta=\sqrt 2$ (union of circles),
$\beta=1$ (single circle), and $\beta=1/\sqrt 2$ (intersection of circles).

As noted above, $\beta$-skeletons were originally introduced for
analyzing empirical networks.
Gabriel graphs and $\beta$-skeletons have many other applications in
{\em computational morphology} (combinatorial methods of representating
shapes).  Gabriel graphs can also be used to construct minimum spanning
trees, since the gabriel graph contains the MST as a subgraph.  More
recently, various researchers have shown that $\beta$-skeletons (for
certain values of $\beta>1$) form subgraphs of the minimum weight
triangulation~\cite{CX95,K94,Y95}.

Su and Chang~\cite{SC90} have described a generalization of Gabriel graphs,
the $k$-Gabriel graphs, in which an edge is present if its diameter circle
contains at most $k-1$ other points.  One can similarly generalize
$\beta$-skeletons to $k$-$\beta$-skeletons.  Our results can be made
to hold as well for these generalizations as for the original graph classes.

\section{Fractals and dilation}

\begin{figure}
$$\efig{3in}{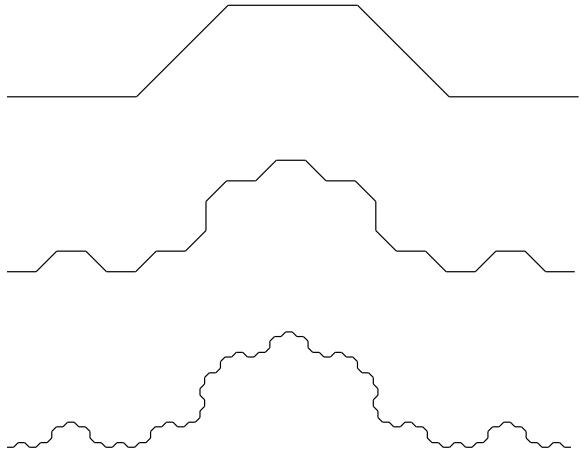}$$
\caption{Fractal curves $P(\pi/4,k)$ for $k=1,2,3$.}
\label{F:fp}
\end{figure}

\begin{figure}
$$\efig{3in}{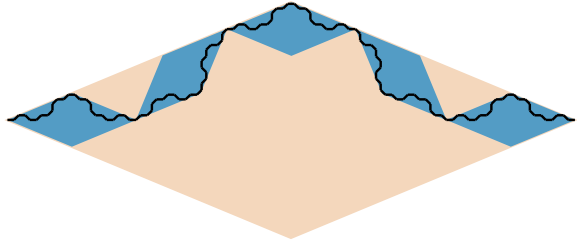}$$
\caption{Fractal curve is contained in a diamond.}
\label{F:fd}
\end{figure}

Our construction showing that beta-skeletons have unbounded dilation
consists of a fractal curve with a recursive definition similar to that
of a Koch snowflake.  For a given angle $\theta$ define the polygonal
path $P(\theta,1)$, by following a path of five equal-length line segments:
one horizontal, one at angle $\theta$, a second horizontal,
a segment at angle $-\theta$, and a third horizontal.

We then more generally define the graph $P(\theta,k)$ to be a path
of $5^k$ line segments, formed by replacing the five segments of $P(\theta,1)$
with congruent copies of $P(\theta,k-1)$, scaled so that the two
endpoints of the path are at distance one from each other.  Figure~\ref{F:fp}
shows three levels of this construction.  In the drawing of Figure~\ref{F:fp},
the orientations of the five copies of $P(\theta,k-1)$ alternate along
the overall path, so that the horizontal copies are in the same orientation
as the overall path and the other two copies are close to upside-down,
but this choice of orientation is not essential to our construction.

Note that, if we denote the length of $P(\theta,k)$ by $\ell_k=\ell_k(\theta)$,
then $\ell_1>1$ and $\ell_k=\ell_1^k$.

\begin{lemma}\label{L:diamond}
$P(\theta,k)$ is contained within a diamond shape having the endpoints of the
path as its diagonal, and with angle $\theta$ at those two corners of
the diamond.
\end{lemma}

\begin{proof}
This follows by induction, as shown in Figure~\ref{F:fd},
since the five such diamonds containing the five copies of $P(\theta,k-1)$
fit within the larger diamond defined by the Lemma.
\end{proof}

\begin{lemma}
If $\theta<(\pi-\sin^{-1}\beta)/2$,
$P(\theta,k)$ is the $\beta$-skeleton of its vertices.
\end{lemma}

\begin{proof}
We show that, if $a$ and $b$ are non-adjacent vertices in the path,
then there is some $c$ forming an angle of at least
$\pi-\sin^{-1}\beta$.
We can assume that $a$ and $b$ are in different copies of $P(\theta,k-1)$,
since otherwise the result would hold by induction.
But no matter where one places two points in different copies of the
small diamonds
containing the copies of $P(\theta,k-1)$ (depicted in Figure~\ref{F:fd}),
we can choose one of the three interior vertices of $P(\theta,1)$
as the third point $c$ forming an angle $acb \ge \pi-2\theta$.
The result follows from the assumed inequality relating $\theta$ to $\beta$.
\end{proof}

For instance, the graphs $P(\pi/4,k)$ depicted in Figure~\ref{F:fp}
are Gabriel graphs of their vertices.  A more careful analysis shows that
larger values of $\theta$ still result in a $\beta$-skeleton:
if the orientations of the copies of $P(\theta,k-1)$ that form $P(\theta,k)$ 
are chosen carefully, $P(\theta,k)$  is contained in only half the diamond
of Lemma~\ref{L:diamond}, and angle $acb$ in the proof above
can be shown to be $\ge \pi-3\theta/2$.

\begin{theorem}
For any $\beta>0$ there is a $c>0$ such that $\beta$-skeletons
of $n$-point sets have dilation $\Omega(n^c)$.
\end{theorem}

\begin{proof}
We have seen that we can choose a $\theta$ such
that the graphs $P(\theta,k)$ are $\beta$-skeletons.
Since the endpoints of the path are at distance one from each other,
the dilation of $P(\theta,k)$ is $\ell_k=\ell_1^k$.
Each such graph has $n=5^k+1$ vertices
and dilation $\ell_1^k=n^{\log_5\ell_1-o(1)}$.
Since $\ell_1>1$, $\log_5\ell_1>0$.
\end{proof}	

\section{Upper Bounds}

We have shown a lower bound of $\Omega(n^c)$ for the dilation of $\beta$-skeletons,
where $c$ is a constant depending on $\beta$, and approaching zero as
$\beta$ approaches zero.  This behavior of having length a fractional power
of $n$ is characteristic of fractal curves; is it inherent in $\beta$-skeletons
or an artifact of our fractal construction?  We now show the former
by proving an upper bound on dilation of the same form.

To do this, we define an algorithm for finding short paths in
$\beta$-skeletons.  As a first start towards such an algorithm,
we use the following simple recursion: to 
find a path from $s$ to $t$, test whether edge $st$ exists in the
$\beta$-skeleton.  If so, use that edge as path.  If not, some $r$
forms a large angle $srt$; concatenate the results of recursively finding
paths from $s$ to $r$ and $r$ to $t$.

\begin{figure}
$$\efig{3in}{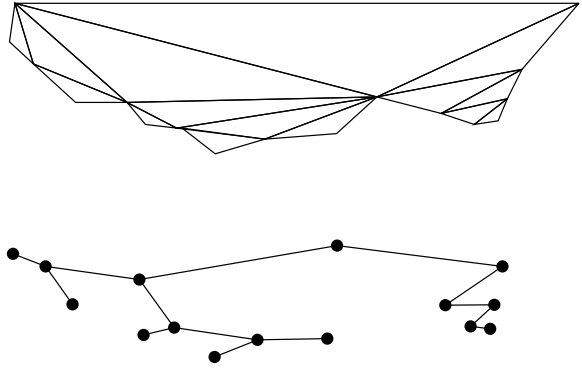}$$
\caption{Tree of triangles and corresponding abstract tree.}
\label{F:tt}
\end{figure}

For $\beta\le 1$, $sr$ and $rt$ are shorter than $st$, so this algorithm
always terminates; we assume throughout the rest of the section that
$\beta\le 1$.  We can represent the path it finds as a tree of triangles, all
having an angle of at least
$\pi-\sin^{-1}\beta$, rooted at triangle $srt$ (Figure~\ref{F:tt}).  The
hypotenuse of each triangle in this tree is equal to one of the two shorter
sides of its parent.  Note that the triangles may overlap geometrically, or
even coincide; we avoid complications arising from this possibility by only
using the figure's combinatorial tree structure.  We will bound the length of
the path found by this algorithm by manipulating trees of this form.  For any
similarly defined tree of triangles, we define the {\em boundary length} of the
tree to be the following formula:
$$|T| = {\rm dist}(s,t)
+ \sum_{\Delta\in T} ({\rm perim}(\Delta) - 2\cdot{\rm hypotenuse}(\Delta)).$$
In other words, we sum the lengths of all
the short sides of the triangles, and subtract the lengths of all non-root
hypotenuses.  If the tree forms a non-self-intersecting polygon, such as the
one shown in the figure, this is distance from $s$ to $t$ ``the long way''
around the polygon's perimeter 

\begin{lemma}
For the tree defined by the algorithm above, $|T|$ is the length of the
path constructed by the algorithm.
\end{lemma}

\begin{proof}
This can be shown by induction using the fact that
the path from $s$ to $t$ is formed by concatenating
the paths from $s$ to $r$ and $r$ to $t$.
\end{proof}

Our bound will depend on the number of leaves in the tree produced above. 
However, this number may be very large, larger than $n$, because the same
vertex of our input point set may be involved in triangles in many unrelated
parts of the tree.  Our first step is to prune the tree to produce one that
still corresponds in a sense to a path in the $\beta$-skeleton, but with a
good bound on the number of leaves.

\begin{lemma}\label{L:pathtree}
For any $\beta\le 1$, we can find a tree like the one described above, with at
most $2n$ leaves, for which $|T|$ is the length of some path in the
$\beta$-skeleton from $s$ to $t$.
\end{lemma}

\begin{proof}
Define a ``leaf vertex'' to be the vertex opposite the hypotenuse of a leaf
triangle in $T$.
We prune the tree one step at a time until each
vertex appears at most twice as a leaf vertex.  
At each step, the path corresponding to
$T$ (and with length at most $|T|$) will visit all the leaf vertices in tree
order (as well as possibly visiting some other vertices coming from interior
nodes of the tree).

Suppose some vertex $v$ appears three or more times.
Then we prune $T$ by removing all subtrees
descending from the path between
the first and last appearance of $v$ (occurring between the two appearences in
tree order), and we shorten the corresponding path by removing the portion of it
between these two appearances of $v$. At each step, the change to $|T|$ comes
from subtracting some triangle short side lengths corresponding to the subtrees
removed from
$T$, as well as adding some hypotenuses of triangles from the same subtrees. 
Each subtracted side length that is not cancelled by an added hypotenuse
corresponds to one of the edges removed from the path, so the total reduction
in $|T|$ is at most as great as the total reduction in the length of the path,
and the invariant that $|T|$ bounds the path length is maintained.
After this pruning, there will be no leaves between the two appearances of
$v$, and no new leaves are created elsewhere in the tree, so the invariant that
the path visits the leaf vertices in order is also maintained.

This pruning process removes at least one appearance of $v$, and so can be
repeated at most finitely many times before terminating.
\end{proof}

\begin{figure}
$$\efig{3in}{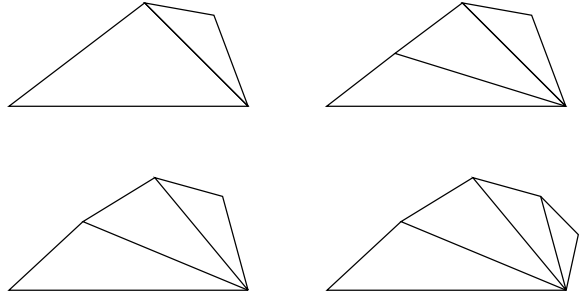}$$
\caption{Making single-leaf tree longer: subdivide triangles, increase
angles, add children.}
\label{F:sz}
\end{figure}

We use induction on the number of leaves to prove bounds on $|T|$.
The following lemma forms the base case:

\begin{lemma}\label{L:oneleaf}
Let $T$ be a tree of triangles, all having an angle of at least $\theta>\pi/2$
opposite the edge connecting to the parent in the tree, with exactly one
leaf triangle, and scaled so that the hypotenuse of the root triangle has
length~$1$.  Then $|T|\le -1/\cos\theta$.
\end{lemma}

\begin{figure}
$$\efig{3in}{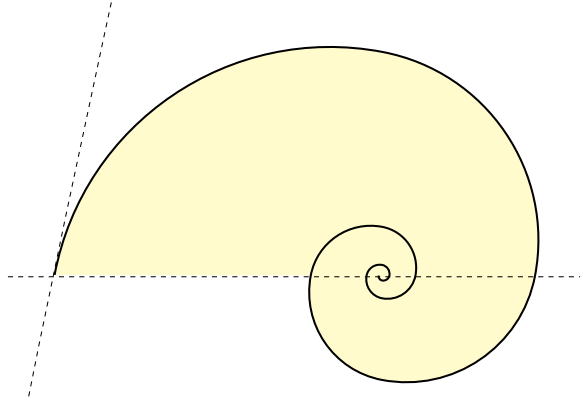}$$
\caption{Logarithmic spiral formed by keeping fixed angle to destination point.}
\label{F:ls}
\end{figure}

\begin{proof}
Since $|T|$ does not depend on the ordering of tree nodes, we can assume
without loss of generality that each node's child is on the left.
For any such tree, we can increase $|T|$ by performing a sequence of the
following steps: (1) If any triangle has an angle greater than $\theta$,
change it to one having an angle exactly equal to $\theta$, without changing
any other triangle shapes.  (2) If any triangle has a ratio of left to right
side lengths less than some value $C$, split it into two triangles by adding a
vertex on the right side.  (3) Add a child to the leaf of $T$.  These steps
are depicted in Figure~\ref{F:sz}.

The result of this sequence of transformations is the concatenation of many
triangles with angles equal to $\theta$, very short left sides, and right sides
with length close to that of the hypotenuse.  In the limit we get a curve from
$s$ to $t$ formed by moving in a direction forming an angle $\pi-\theta$ to $t$,
namely the {\em logarithmic spiral} (Figure~\ref{F:ls}). Integrating the
distance traveled on this spiral against the amount by which the distance to
$t$ is reduced shows that it has the length formula claimed in the lemma.
Since we reach this limit by a monotonically increasing sequence of tree
lengths, starting with any finite one-leaf tree, any finite tree must have
length less than this limit.
\end{proof}

More generally, we have the following result.

\begin{lemma}\label{L:treelength}
Let $T$ be a tree of triangles, all having an angle of at least $\theta>\pi/2$
opposite the edge connecting to the parent in the tree, with $k$
leaf triangles, and scaled so that the hypotenuse of the root triangle has
length~$1$.  Then $|T|\le (-1/\cos\theta)^{1+\lfloor\log_2 k\rfloor}$.
\end{lemma}

\begin{proof}
We prove the result by induction on $k$; Lemma~\ref{L:oneleaf} forms the base
case.  If there is more than one leaf in $T$, form a smaller tree $T'$ by
removing from $T$ each path from a leaf to the nearest ancestor with more than
one child.  These paths are disjoint, and each such removal replaces a subtree
with one leaf by the edge at the root of the subtree, so using
Lemma~\ref{L:oneleaf} again shows that $|T|\le -|T'|/\cos\theta$.
Each leaf in $T'$ has two leaf descendants in $T$, so the number of leaves in
$T$ is at most $k/2$ and the result follows.
\end{proof}

This, finally, provides a bound on $\beta$-skeleton dilation.

\begin{figure}
$$\efig{3in}{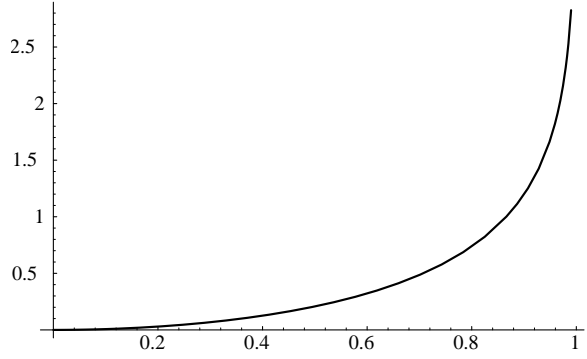}$$
\caption{Exponent in the bound of Theorem~\protect\ref{T:ub}, as a function of
$\beta$.}
\label{F:exp}
\end{figure}

\begin{theorem}\label{T:ub}
For $\beta<\sqrt{3}/2\approx 0.866025$, any $\beta$-skeleton has dilation
$O(n^c)$, where $c<1$ is a constant depending on $\beta$ and going to zero in
the limit as $\beta$ goes to zero.
\end{theorem}

\begin{proof}
We have seen (Lemma~\ref{L:pathtree}) that we can connect any pair of vertices
in the skeleton by a path with length bounded by $|T|$, where $T$ is a tree of
triangles in which all angles are at least $\pi-\sin^{-1}\beta$, and where $T$
has at most $2n$ leaves.  By Lemma~\ref{L:treelength},
the length of such a tree is at most
$$
(-1/\cos(\pi-\sin^{-1}\beta))^{1+\lfloor\log_2 2n\rfloor}=
O\big(n^{\log_2{{\scriptstyle -1}\over
{\scriptstyle\cos(\pi-\sin^{-1}\beta)}}}\big)=
O\big(n^{-{1\over 2}\log_2(1-\beta^2)}\big)
$$
which has the form specified in the statement of
the theorem.
\end{proof}

Figure~\ref{F:exp} shows the growth of the exponent $c$ as a function of
$\beta$.  For $\sqrt 3 / 2 \le \beta \le 1$,
the theorem does not give the best bounds; a bound of $n-1$ on dilation can be
proven using the fact that the skeleton contains the minimum spanning tree.

\section*{Acknowledgements}

Work supported in part by NSF grant
CCR-9258355 and by matching funds from Xerox Corp.
Thanks to Marshall Bern for suggesting the problem
of $\beta$-skeleton dilation.

\end{document}